\documentclass{ere04}

\usepackage{graphicx}

\begin{document}

\title*{Linearized warp drive and the energy conditions}
\author{Francisco S. N. Lobo\inst{1}\and
Matt Visser\inst{2}}
\institute{Centro de Astronomia e Astrof\'{\i}sica da Universidade
de Lisboa, \\Campo Grande, Ed. C8 1749-016 Lisboa, Portugal.
\texttt{flobo@cosmo.fis.fc.ul.pt} \and School of Mathematics,
Statistics, and Computer Science, Victoria University of
Wellington, \\ P.O.Box 600, Wellington, New Zealand.
\texttt{matt.visser@mcs.vuw.ac.nz}}
%
%
\maketitle

``Warp drive'' spacetimes are useful as ``gedanken-experiments''
and as a theoretician's probe of the foundations of general
relativity. Applying linearized gravity to the weak-field warp
drive, i.e., for non-relativistic warp-bubble velocities, we find
that the occurrence of energy condition violations in this class
of spacetimes is generic to the form of the geometry under
consideration and is not simply a side-effect of the
``superluminal'' properties. Using the linearized construction it
is now possible to compare the warp field energy with the
mass-energy of the spaceship, and applying the ``volume integral
quantifier'', extremely stringent conditions on the warp drive
spacetime are found.

\section{Introduction}
\label{sec:1}

``Warp drive'' spacetimes~\cite{Alcubierre,Natario} are specific
examples of solving the Einstein field equation in the reverse
direction, in which one engineers an interesting spacetime metric,
then finds the matter distribution responsible for the respective
geometry. The analysis of wormhole geometries is treated in a
similar manner~\cite{Morris,Visser}. Following this philosophy, it
was found that these spacetimes violate the energy conditions of
general relativity. Although most (but not all) classical forms of
matter are thought to obey the energy conditions, they are
definitely violated by certain quantum fields~\cite{twilight}.
Another interesting feature of these spacetimes is that they allow
``effective'' superluminal travel, although, locally, the speed of
light is not surpassed~\cite{LoboSLT}. However, to provide a
general \emph{global} definition of superluminal travel is no
trivial matter~\cite{VBL}. Nevertheless, it was found that
negative energy densities and superluminal travel are intimately
related~\cite{Olum}, and that certain classical systems, such as
non-minimally coupled scalar fields, violate the null and the weak
energy conditions~\cite{B&V,BV1}. Another severe drawback is that
by using the ``quantum inequality''~\cite{F&R1} it can be argued
that truly enormous amounts of energy are needed to sustain
superluminal spacetimes~\cite{Roman,PfenningF,Everett}.

In this work we will not focus our attention on the superluminal
features of the ``warp drive''~\cite{LV}, but rather on the
weak-field limit, considering the bubble velocity to be
non-relativistic, $v\ll 1$. We shall be interested in applying
linearized gravity to warp drive spacetimes, testing the energy
conditions at first and second order of the warp-bubble velocity.
A particularly interesting aspect of this construction is that one
may now place a finite mass spaceship at the origin and
consequently compare the mass-energy of the warp field with the
mass-energy of the spaceship. This is not possible in the usual
finite-strength warp field, since in the usual formalism the
spaceship is always treated as a massless test particle, moving
along a geodesic. It is interesting to note that if it is possible
to realise even a weak-field warp drive in nature, such a
spacetime appears to be an example of a ``reaction-less drive''.
That is, the warp bubble moves by interacting with the geometry of
spacetime instead of expending reaction mass, and the spaceship is
simply carried along with it.

We shall also apply the ``volume integral quantifier'', as defined
in \cite{Kar1,Kar2}, to the weak-field limit, at first and second
order of the warp-bubble velocity, and thus, find extremely
stringent conditions on the warp drive spacetime. If the
construction of a ``strong-field'' warp drive starts from an
approximately Minkowski spacetime, and inexorably passes through a
weak-field regime, these conditions are so stringent that it
appears unlikely that the ``warp drive'' will ever prove
technologically useful. However, ``warp drive'' spacetimes are
likely to retain their status as useful ``gedanken-experiments'',
as they are useful primarily as a theoretician's probe of the
foundations of general relativity.

\section{Alcubierre warp drive}

Alcubierre demonstrated, within the framework of general
relativity, that it is in principle possible to warp spacetime in
a small {\it bubble-like} region, in such a way that the bubble
may attain arbitrarily large velocities~\cite{Alcubierre}. The
enormous speed of separation arises from the expansion of
spacetime itself. The Alcubierre model for hyper-fast travel
resides in creating a local distortion of spacetime, producing an
expansion behind the bubble, and an opposite contraction ahead of
it.

The warp drive spacetime metric, in cartesian coordinates, is
given by (with $G=c=1$)
\begin{eqnarray}
d s^2=-d t^2+ [d\vec x - \vec \beta(x,y,z-z_0(t)) \; d t]
      \cdot [d\vec x - \vec \beta(x,y,z-z_0(t)) \; d t]\,.
\label{Cartesianwarpmetric-general}
\end{eqnarray}
In terms of the ADM formalism, this corresponds to a spacetime
wherein \emph{space} is flat, while the ``lapse function'' is
identically unity, and the only non-trivial structure lies in the
``shift vector'' $\beta(t,\vec x)$. The Alcubierre warp drive
corresponds to taking the shift vector to lie in the direction of
motion, i.e.,
$\vec \beta(x,y,z-z_0(t)) =  v(t) \; \hat z \; f(x,y,z-z_0(t))$,
%
in which $v(t)=dz_0(t)/dt$ is the velocity of the warp bubble,
moving along the positive $z$-axis. An alternative to the
Alcubierre spacetime, is the Nat\'{a}rio warp drive
\cite{Natario}, where the shift vector is constrained by being
divergence-free, $\nabla \cdot \vec \beta(x,y,z) =  0$.

The form function $f(x,y,z)$ possesses the general features of
having the value $f=0$ in the exterior and $f=1$ in the interior
of the bubble.  The general class of form functions chosen by
Alcubierre was spherically symmetric: $f(r(t))=f(x,y,z-z_0(t))$
with $r(t)=\left\{[(z-z_{0}(t)]^2+x^2+y^2\right\}^{1/2}$. Whenever
a more specific example is required we adopt
\begin{equation}
f(r)=\frac{\tanh\left[\sigma(r+R)\right]
-\tanh\left[\sigma(r-R)\right]}{2\tanh(\sigma R)}\,,
\label{E:form}
\end{equation}
in which $R>0$ and $\sigma>0$ are two arbitrary parameters. $R$ is
the ``radius'' of the warp-bubble, and $\sigma$ can be interpreted
as being inversely proportional to the bubble wall thickness.

Note that observers with the four velocity
$U^{\mu}=\left(1,0,0,vf\right)$ move along geodesics, as their
$4$-acceleration is zero, \emph{i.e.}, $a^{\mu} = U^{\nu}\,
U^{\mu}{}_{;\nu}=0$. The hypothetical spaceship, which in the
original formulation is treated as a test particle and placed
within the Alcubierre warp bubble, moves along a timelike curve
$z=z_0(t)$ regardless of the value of $v(t)$. Proper time along
this curve equals the coordinate time~\cite{Alcubierre} and the
centre of the perturbation corresponds to the spaceship's position
$z_0(t)$. The expansion of the volume elements,
$\theta=U^{\mu}{}_{;\mu}$, is given by $\theta=v\;\left({\partial
f}/{\partial z} \right)$. Using equation (\ref{E:form}), one may
prove
that the volume elements are expanding behind the spaceship, and
contracting in front of it \cite{Alcubierre,LV}.

If we treat the spaceship as more than a test particle, we must
confront the fact that by construction we have forced $f=0$
outside the warp bubble. This implies that the spacetime geometry
is asymptotically Minkowski space, and in particular the ADM mass
is zero. That is, the ADM mass of the spaceship and the warp field
generators must be exactly compensated by the ADM mass due to the
stress-energy of the warp-field itself. Viewed in this light it is
now patently obvious that there must be massive violations of the
classical energy conditions. One of our tasks in the current
article will be to see if we can make qualitative and quantitative
statements concerning the localization and ``total amount'' of
energy condition violations. A similar attempt at quantification
of the ``total amount'' of energy condition violation in
traversable wormholes was recently presented in~\cite{Kar1,Kar2},
by introducing the notion of the ``volume integral quantifier''.
This notion amounts to calculating the definite integrals $\int
T_{\mu\nu}U^\mu U^\nu \,dV$ and $\int T_{\mu\nu}k^\mu k^\nu \,dV$,
and the amount of violation is defined as the extent to which
these integrals become negative, where $U^{\mu}$ and $k^\mu$ are
any timelike and null vectors, respectively, and $T_{\mu\nu}$ is
the stress-energy tensor.

The weak energy condition (WEC) states $T_{\mu\nu}\, U^{\mu}
U^{\nu}\geq0$. Its physical interpretation is that the local
energy density is positive.  By continuity it implies the null
energy condition (NEC). We verify that for the warp drive metric,
the WEC is violated, \emph{i.e.},
\begin{equation}\label{WECviolation}
T_{\mu\nu} \, U^{\mu} \, U^{\nu}= -\frac{v^2}{32\pi}\; \left[
\left (\frac {\partial f}{\partial x} \right )^2 + \left (\frac
{\partial f}{\partial y} \right )^2 \right] =
-\frac{1}{32\pi}\frac{v^2 (x^2+y^2)}{r^2} \left( \frac{d f}{d r}
\right)^2 <0 \,.
\end{equation}

Using an orthonormal basis, the energy density of the warp drive
spacetime is given by $T_{\hat{t}\hat{t}} = T_{\hat{\mu}\hat{\nu}}
\, U^{\hat{\mu}} \, U^{\hat{\nu}}$, i.e., eq.
(\ref{WECviolation}), which is distributed in a toroidal region
around the $z$-axis, in the direction of travel of the warp
bubble~\cite{PfenningF}. Note that the energy density for this
class of spacetimes is nowhere positive, and the fact that the
total ADM mass can nevertheless be zero is due to the intrinsic
nonlinearity of the Einstein equations.

We can, in analogy with the definitions in~\cite{Kar1,Kar2},
quantify the ``total amount'' of energy condition violating matter
in the warp bubble by defining
\begin{eqnarray}
M_\mathrm{warp} = \int \rho_\mathrm{warp} \; d^3 x = \int
T_{\mu\nu} \, U^{\mu}\, U^{\nu}  \; d^3 x
     = -{v^2\over12} \int \left( \frac{d f}{d r} \right)^2 \; r^2 \;
d r.
\end{eqnarray}
This is emphatically not the total mass of the spacetime, but it
characterizes the amount of negative energy that one needs to
localize in the walls of the warp bubble. For the specific shape
function (\ref{E:form}) we can estimate $M_\mathrm{warp} \approx -
v^2 \, R^2 \, \sigma$.

The NEC states that $T_{\mu\nu} \, k^{\mu} \, k^{\nu}\geq0$.
Considering the NEC for a null vector oriented along the $\pm \hat
z$ directions, and in particular, if we average over the $\pm \hat
z$ directions we have~\cite{LV}
\begin{eqnarray}
{1\over2} \left\{ T_{\mu\nu} \, k^{\mu}_{+\hat z} \,
k^{\nu}_{+\hat z} + T_{\mu\nu} \, k^{\mu}_{-\hat z} \,
k^{\nu}_{-\hat z} \right\} = -\frac{v^2}{8\pi}\, \left[ \left
(\frac{\partial f}{\partial x} \right )^2 + \left (\frac {\partial
f}{\partial y} \right )^2 \right],
\end{eqnarray}
which is manifestly negative, and so the NEC is violated for all
$v$.

Using the ``volume integral quantifier'', we may estimate the
``total amount'' of averaged null energy condition violating
matter in this spacetime, given by
$\int T_{\mu\nu} \,k^{\mu}_{\pm\hat z} \, k^{\nu}_{\pm\hat z}  \;
d^3 x   \approx - v^2\, R^2 \; \sigma \approx M_{\mathrm{warp}}$.
%
The key aspects to note here are that the net volume integral of
the $O(v)$ term is zero, and that the net volume average of the
NEC violations is approximately the same as the net volume average
of the WEC violations, which are $O(v^2)$~\cite{LV}.

\section{Linearized warp drive}

Our goal now  is to try to build a more realistic model of a warp
drive spacetime where the warp bubble is interacting with a finite
mass spaceship. To do so we first consider the linearized theory
applied to warp drive spacetimes, for non-relativistic velocities,
$v\ll 1$. In linearized theory, the spacetime metric is given by
$d s^2=\left(\eta_{\mu\nu} + h_{\mu\nu}\right)\, d x^\mu \, d
x^\nu$, with $h_{\mu\nu}\ll 1$ and $\eta_{\mu\nu}={\rm
diag}(-1,1,1,1)$. The analysis can be simplified by defining the
{\it trace reverse} of $h_{\alpha\beta}$, given by
$\overline{h}_{\alpha\beta} =
h_{\alpha\beta}-\frac{1}{2}\eta_{\alpha\beta}\,h
\label{tracereverse}$,
with $\overline{h}=\overline{h}^{\alpha}{}_{\alpha}=-h$. In terms
of $\overline{h}_{\alpha\beta}$, the linearized Einstein tensor
reads
\begin{eqnarray}
G_{\alpha\beta} = -
\frac{1}{2}\Big[\overline{h}_{\alpha\beta,\mu}{}^{\mu} +
\eta_{\alpha\beta} \overline{h}_{\mu\nu,}{}^{\mu\nu} -
\overline{h}_{\alpha\mu,\beta}{}^{\mu}
    - \overline{h}_{\beta\mu,\alpha}{}^{\mu} +
O\left(\overline{h}_{\alpha\beta}^2\right)\Big]
\label{linearEinstein} \,.
\end{eqnarray}

Now the results deduced from applying linearized theory are only
accurate to first order in $v$. This is equivalent to neglecting
the $v^2f^2$ from the metric (\ref{Cartesianwarpmetric-general}),
retaining only the first order terms in $v$. That is, we are
making the following approximation
\begin{equation}
(h_{\mu\nu})=\left[
\begin{array}{cccc}
0&0&0&-vf \\
0&0&0&0 \\
0&0&0&0 \\
-vf&0&0&0
\end{array}
\right] \label{linearperturbation}\,.
\end{equation}
The trace of $h_{\mu\nu}$ is identically null, \emph{i.e.},
$h=h^{\mu}{}_{\mu}=0$. Therefore, the trace reverse of
$h_{\mu\nu}$, defined in eq. (\ref{tracereverse}), is given by
$\overline{h}_{\mu\nu}=h_{\mu\nu}$, i.e., eq.
(\ref{linearperturbation}) itself.

\subsection{Energy condition violations}

In linearized theory the $4$-velocity can be approximated by
$U^{\mu}=(1,0,0,0)$, and by using equation (\ref{linearEinstein})
we verify that the WEC is identically ``saturated'', i.e.,
$T_{\mu\nu} \, U^{\mu} \, U^{\nu}=T_{00}=O(v^2)$.
%
%
Although in this approximation, at least to first order in $v$,
the WEC is not violated, it is on the verge of being so.

Despite the fact that the observers, with $U^{\mu}=(1,0,0,0)$,
measure zero energy density, it can be shown that observers which
move with any other arbitrary velocity, $\tilde\beta$, along the
positive $z$ axis measure a negative energy density, to first
order in $v$. Consider a Lorentz transformation, and using
equation (\ref{linearEinstein}) the energy density measured by
these observers is given by
\begin{eqnarray}
T_{\hat{0}\hat{0}}= \frac{\gamma^{2}\tilde\beta v}{8\pi} \Bigg[
\left(\frac{x^2+y^2}{r^2}\right) \; \frac{d^2 f}{d r^2} +
\left(\frac{x^2+y^2+2(z-z_0(t))^2}{r^3}\right) \; \frac{d f}{d r}
\Bigg]  + O(v^2) \label{negativeenergydensity}\,.
\end{eqnarray}
with $\gamma=(1-\tilde\beta)^{-1/2}$. A number of general features
can be extracted from the terms in square brackets, without
specifying an explicit form of $f$. In particular, $f$ decreases
monotonically from its value at $r=0$, $f=1$, to $f\approx 0$ at
$r\geq R$, so that ${d f}/{d r}$ is negative in this domain. The
form function attains its maximum in the interior of the bubble
wall, so that ${d^2 f}/{d r^2}$ is also negative in this region.
Therefore there is a range of $r$ in the immediate interior
neighbourhood of the bubble wall where one necessarily encounters
negative energy density, as measured by the observers considered
above. Again we find that WEC violations persist to arbitrarily
low warp bubble velocities.

One can show that the NEC is proportional to the energy density,
$T_{\hat{0}\hat{0}}$, of equation (\ref{negativeenergydensity}).
Thus, we verify that the NEC is also violated in the immediate
interior vicinity of the bubble wall (see \cite{LV} for details).

\subsection{Spaceship immersed in the warp bubble}

Consider now a spaceship in the interior of an Alcubierre warp
bubble, which is moving along the positive $z$ axis with a
non-relativistic constant velocity, i.e., $v\ll 1$. The metric is
given by
\begin{eqnarray}
&\hspace{-2.5cm}d s^2=-d t^2+d x^2+d y^2+\left[d
z-v\;f(x,y,z-vt)\,d t \right]^2
\nonumber   \\
&-2\Phi(x,y,z-vt)\, \big[d t^2+d x^2+d y^2
   +(d z-v\;f(x,y,z-vt)\,d t)^2 \big] \label{warpspaceshipmetric}
\,.
\end{eqnarray}
If $\Phi =0$, the metric (\ref{warpspaceshipmetric}) reduces to
the warp drive spacetime of eq.
(\ref{Cartesianwarpmetric-general}). If $v=0$, we have the metric
representing the gravitational field of a static source, in
particular, that of a spaceship. Note that the mass density of the
spaceship, $\rho$, is related to the gravitational potential
$\Phi$ by Poisson's equation, $\nabla ^2\Phi =4\pi \rho$.

\medskip

\emph{First order approximation}

Applying the linearized theory, keeping terms linear in $v$ and
$\Phi$ but neglecting all superior order terms, the matrix
elements, $h_{\mu\nu}$, and the respective trace-reversed
elements, $\overline{h}_{\mu\nu}$, of the metric
(\ref{warpspaceshipmetric}) are given by the following
approximations
\begin{equation}
(h_{\mu\nu})=\left[
\begin{array}{cccc}
-2\Phi&0&0&-vf \\
0&-2\Phi&0&0 \\
0&0&-2\Phi&0 \\
-vf&0&0&-2\Phi
\end{array}
\right]
\qquad \,{\rm and} \,\qquad
(\overline{h}_{\mu\nu})=\left[
\begin{array}{cccc}
-4\Phi&0&0&-vf \\
0&0&0&0 \\
0&0&0&0 \\
-vf&0&0&0
\end{array}
\right]   \,,
\end{equation}
where the trace of $h_{\mu\nu}$ is given by
$h=h^{\mu}_{\;\;\mu}=-4\Phi$.

We verify that the WEC is given by $T_{\mu\nu}U^{\mu}U^{\nu}=\rho
+ O(v^2, v \Phi, \Phi^2)$,
%
%
where $\rho$ is the energy density of the spaceship which is
manifestly positive.  In linearized theory, the total ADM mass of
the space-time simply reduces to the mass of the space-ship, i.e.,
\begin{eqnarray}
M_\mathrm{ADM}=\int T_{00} \, d ^3 x=\int \rho \; d ^3 x  + O(v^2,
v \Phi, \Phi^2)
        = M_\mathrm{ship} + O(v^2, v \Phi, \Phi^2)\,.
\end{eqnarray}

The NEC, with $k^{\mu}\equiv (1,0,0,\pm 1)$, to first order in $v$
and $\Phi$, and neglecting the crossed terms $v \, \Phi$, takes
the form
\begin{equation}
T_{\mu\nu} k^{\mu} k^{\nu}=\rho
\pm\frac{v}{8\pi}\left(\frac{\partial^2 f}{\partial
x^2}+\frac{\partial^2 f}{\partial y^2}\right) + O(v^2, v \Phi,
\Phi^2) \,.
\end{equation}
From this, one can deduce the existence of localized NEC
violations even in the presence of a finite mass spaceship, and
can also make deductions about the net volume-averaged NEC
violations. First, note that for reasons of structural integrity
one wants the spaceship itself to lie well inside the warp bubble,
and not overlap with the walls of the warp bubble.  But this means
that the region where $\rho\neq0$ does not overlap with the region
where the $O(v)$ contribution due to the warp field is non-zero.
So regardless of how massive the spaceship itself is, there will
be regions in the wall of the warp bubble where localized
violations of NEC certainly occur.  If we now look at the volume
integral of the NEC, we have
\begin{eqnarray}
\int T_{\mu\nu} \;k^{\mu}_{\pm\hat z} \; k^{\nu}_{\pm\hat z}  \;
d^3 x = \int \rho \; d ^3 x  + O(v^2,v\Phi,\Phi^2)
   =M_\mathrm{ship} + O(v^2,v\Phi,\Phi^2)\,.
\end{eqnarray}
The net result of this $O(v)$ calculation is that irrespective of
the mass of the spaceship there will always be localized NEC
violations in the wall of the warp bubble, and these localized NEC
violations persist to arbitrarily low warp velocity. However at
$O(v)$ the volume integral of the NEC violations is zero, and so
we must look at higher order in $v$ if we wish to deduce anything
from the consideration of volume integrals to probe ``net''
violations of the NEC.

\medskip

\emph{Second order approximation}


Consider the approximation in which we keep the exact $v$
dependence but linearize in the gravitational field of the
spaceship $\Phi$. The WEC is given by (see \cite{LV} for details)
\begin{eqnarray}
T_{\hat{\mu}\hat{\nu}} \, U^{\hat{\mu}} \, U^{\hat{\nu}}=\rho -
\frac{v^2}{32\pi}\left[\left(\frac{\partial f}{\partial
x}\right)^2 + \left(\frac{\partial f}{\partial y}\right)^2\right]
+O(\Phi^2)   \,.
\end{eqnarray}

Once again, using the ``volume integral quantifier'', we find the
following estimate
\begin{eqnarray}
\int T_{\hat{\mu}\hat{\nu}} \, U^{\hat{\mu}} \, U^{\hat{\nu}} \;
d^3 x =  M_{\rm ship} -v^2 \,R^2\,\sigma + \int O(\Phi^2)  \;d^3 x
 \,.
\end{eqnarray}
Now suppose we demand that the volume integral of the WEC at least
be positive, then $v^2 \,R^2\,\sigma  \leq M_{\rm ship}$.
%
%
This inequality is the reasonable condition that the net total
energy stored in the warp field be less than the total mass-energy
of the spaceship itself, which places a powerful constraint on the
velocity of the warp bubble. Re-writing this in terms of the size
of the spaceship $R_\mathrm{ship}$ and the thickness of the warp
bubble walls $\Delta = 1/\sigma$, we have
\begin{equation}\label{restriction}
v^2 \leq {M_{\rm ship}\over R_\mathrm{ship}}\; {R_\mathrm{ship} \;
\Delta\over R^2}.
\end{equation}
For any reasonable spaceship this gives extremely low bounds on
the warp bubble velocity.
In a similar manner, we find the same restriction as eq.
(\ref{restriction}) whilst analyzing the NEC~\cite{LV}.

\section{Conclusion}

We have verified that the warp drive spacetimes necessarily
violate the classical energy conditions, and continue to do so for
arbitrarily low warp bubble velocity. Thus the energy condition
violations in this class of spacetimes is generic to the form of
the geometry under consideration and is not simply a side-effect
of the ``superluminal''  properties.

Using linearized theory, we have built a more realistic model of
the warp drive spacetime where the warp bubble interacts with a
finite mass spaceship. We have applied the ``volume integral
quantifier'' to the weak-field limit, and found that this places
an extremely stringent condition on the warp drive spacetime,
namely, that for all conceivably interesting situations the bubble
velocity should be absurdly low. In view of this analysis, it
therefore appears unlikely that the warp drive will ever prove to
be technologically useful.



\end{document}